\documentclass[twocolumn,showpacs,preprintnumbers,amsmath,amssymb]{revtex4}

\usepackage{graphicx}
\usepackage{dcolumn}
\usepackage{bm}

\begin{document}


\title{Coherent $\rho^0$ photoproduction in bulk matter at high energies}
\author{Elsa Couderc} \affiliation{Lawrence
Berkeley National Laboratory, Berkeley CA 94720 USA \\ and D\'epartement de physique, 
Ecole Normale Sup\'erieure, Paris, France.} 
\author{Spencer Klein} \affiliation{Lawrence Berkeley National Laboratory, Berkeley CA, 94720 
USA \\ and Department of Physics, University of California, Berkeley, CA, 94720 USA}
 
\date{\today}

\begin{abstract}

The momentum transfer $\Delta k$ required for a photon to scatter from a target 
and emerge as a $\rho^0$ decreases as the photon energy $k$ rises.  For $k>3\times10^{14}$ eV, 
$\Delta k$ is small enough that the interaction  cannot be localized to a single nucleus.  At still higher 
energies, photons may coherently scatter elastically from bulk matter and
emerge as a $\rho^0$, in a manner akin to kaon regeneration.  Constructive interference from the different
nuclei coherently raises the cross section and the interaction 
probability rises linearly with energy.  At energies above $10^{23}$ eV, coherent conversion is
the dominant process; photons interact predominantly as $\rho^0$.
We compute the coherent scattering
probabilities in slabs of lead, water and rock, and discuss the implications of the 
increased hadronic interaction probabilities for photons on ultra-high 
energy shower development.  

\end{abstract}



\maketitle
An understanding of the cross sections for photon interactions 
is important in many areas of physics.  For example, several
groups have searched for radio waves \cite{radio} or acoustic pulses \cite{acoustic} produced by interactions of 
astrophysical $\nu_e$ with energies up to $10^{25}$ eV. 
The radio and acoustic frequency spectra and angular distribution depend on the distribution of moving electric charges (for radio) and energy deposition (for acoustic).
These distributions are controlled by the particle interactions that govern 
shower development.   The experimental flux limits depend on the assumed character of these interactions.  

In this letter, we  discuss hadronic interactions of photons \cite{eespencer}, and introduce a new effect: 
coherent photon to
$\rho^0$ conversion in bulk matter. These effects become important at energies above $10^{20}$ eV, but 
are not in many current calculations \cite{klein-erice}.
With these effects, photons produce hadronic showers, rather than electromagnetic showers.  Since hadronic showers are much less
affected by the Landau-Pomeranchuk-Migdal (LPM) effect \cite{zas}, they are much more compact; the presence of hadronic interactions alters both the 
frequency spectrum and angular distribution of the electromagnetic and acoustic radiation from neutrino 
induced showers.  Published limits that do not consider these effects may need to be revised. 

Besides it's importance for $\nu$ searches, coherent conversion is very 
interesting in it's own right, as one of a handful of examples where particles interact very differently in 
bulk matter than with isolated atoms; interactions with individual targets are replaced by distributed
interactions which are de-localized over multiple atoms \cite{kansas}.  The other prominent examples are LPM suppression of bremsstrahlung and
pair production \cite{lpm,stdrock}, kaon regeneration \cite{kaonregeneration}, and coherent neutrino forward scattering \cite{liu}.  
The LPM  effect decreases the cross section for pair production, due to destructive interference.  In contrast,
in the other examples, including coherent $\rho$ conversion,  constructive interference raises the cross section.

Photon to $\rho$ conversion occurs when a photon fluctuates
to a virtual $q\overline q$ pair which then scatters elastically from a target nucleus, 
emerging as a vector meson \cite{bauer}.  As the photon energy $k$ rises,
the required momentum transfer $\Delta k = M_{\pi\pi}^2/2k$ decreases,
and the coherence length $l_f = \hbar/\Delta k$ rises (we take
$\hbar=c=1$ throughout).
Here $M_{\pi\pi}$ is the final state mass.  When $k>3\times10^{14}$ eV 
(for $M_{\pi\pi}= M_\rho =778$ MeV/c$^2$, the $\rho$ pole mass),  $l_f > 0.2$ nm, the
typical internuclear spacing, and coherence over multiple nuclei becomes possible. 

We consider a high-energy plane-wave photon travelling in the $+z$
direction.  The photon wave function is mostly bare photon ($|\gamma_b\!>$),
but with a $q\overline q$ component, $|q\overline q\!>$ and heavier fluctuations
({\it e.g.} $q\overline q q\overline q$, and strange quark pairs): 
\begin{equation}
|\psi> = e^{ikz} \big[\sqrt{1-F^2}|\gamma_b> + F|q\overline q> + ... \big] 
\end{equation}
where $F= 6.02 10^{-2}$ is the amplitude for the photon to fluctuate to a $q \overline q$ 
pair \cite{bauer}.  Heavier fluctuations are not relevant here, and will not be further
considered.

The $|q\overline q\!>$ scatters from nuclei at positions $\vec{r}_i$, with
scattering amplitude $f(\theta)$, emerging with momentum $\vec{k'}$. 
Neglecting photon or $\rho$ absorption, and photon scattering (so $q\overline q$ do not 
scatter and then fluctuate back to a photon), the wavefunction at a depth $L$ 
in the material is 
\begin{equation}
|\!\psi(L)\!>\!=\!\sqrt{1 - F^2}  e^{ikL} |\gamma_b\!>
\!+\!F\!\sum_i {e^{ikz_i} f(\theta) \frac{e^{i\vec{k'_i}(\vec{L}-\vec{r_i})}}{|\vec{L}-\vec{r_i}|} }|\rho\!>.
\end{equation} 

The photon can be absorbed in the target, with absorption
length $\alpha = 1/n[\sigma_{ee}(\gamma A) + \sigma_{hadr}(\gamma A)]$.  Here
$\sigma_{ee}(\gamma A)$ is the pair production cross section,
$\sigma_{hadr}(\gamma A)$ is the photonuclear cross section, and 
$n$ is the target density, in atoms/volume. 

The $\rho$ can be lost by direct hadronic interaction, or by decay to
$\pi^+\pi^-$, followed by $\pi$ interactions.  The decay
itself does not affect the coherence \cite{interfere}, but
two $\pi$ interact differently from one $\rho$.  
For simplicity,  we  take $\beta = 1/n\sigma_{tot}(\rho A)$ 
where $\sigma_{tot}(\rho A)$ is the $\rho$ nucleus cross section. 
The
factor of 2 difference between two $\pi$ and one $\rho$ is small compared to the other uncertainties.
This also applies
for direct $\pi^+\pi^-$ production, discussed below.  

Including these absorption factors, and treating the target as a homogenous medium with constant
density $n$,  the wave function at depth $L$ in the slab is
\begin{eqnarray}
<\rho^0&|&\psi(L)>=nF \int_{0}^{\infty}{\eta d\eta}\int_0^{2\pi}{d\psi} \nonumber \\
&\times & \int_0^L{dz e^{(ik-\frac{\alpha}{2}) z} f(\theta) \frac{e^{(ik'-\frac{\beta}{2})r'}}{r'} },
\end{eqnarray} 
for an incident plane wave moving along the $+z$ axis. 
The integral over the target volume is in cylindrical coordinates where $\eta$ is the
radial distance from the $z$ axis and $\psi$ is the azimuthal angle. We neglect scattering effects at large distances since 
absorption limits the effective size of the target.

In an infinite medium only forward scattering, $f(0)$, contributes to coherent scattering 
\cite{kaonregeneration,liu,lax}. We assume that the real part of $f(0)$ is small \cite{bauer}, 
corresponding to nearly complete absorption.  Then, from the optical theorem,
 $|f(0)| = \sigma_{tot}(\rho A) k/4\pi$.

After scattering, the system has the momentum $k' = k - \Delta k$.
Since $\Delta k \propto M_{\pi\pi}^2$, the $\rho$ width affects the
calculation.  The photon may also fluctuate
directly into an $\pi^+\pi^-$ pair \cite{soeding}; these two processes interfere and the
mass spectrum is a Breit-Wigner spectrum.  Then,

\begin{eqnarray}
f(0,M_{\pi\pi}) = \frac{k}{4\pi}  \bigg| A \frac{\sqrt{M_{\pi \pi} m_{\rho} \Gamma_{\rho}}}{M^2_{\pi \pi} - m^2_{\rho}+i m_{\rho} \Gamma_{\rho}}+B \bigg|^2
\end{eqnarray}
where $A$ and $B$ are the energy-dependent
amplitudes for $\rho$ and direct $\pi\pi$ production
respectively \cite{joakimspencer}, and
the $\rho^0$ width $\Gamma_{\rho} = 150$ MeV.
We assume that $B/A$ is
independent of photon energy and target material.
Integrating $\int dM_{\pi\pi} f(0,M_{\pi\pi})$ returns the traditional $f(0)$.
With this,  $\sigma(\gamma p\rightarrow \rho p) = A^2$.

We substitute $\eta d\eta = r'dr'$ where $r'$ is the distance between the
scattering point $(\eta,z)$ and the observer at $(0,L$),
to evaluate the integrals \cite{liu}:
\begin{eqnarray}
<\rho^0|\psi(L)> =  2 \pi nF \int_{2m_\pi}^{M_{\rho}+5\Gamma_\rho} dM_{\pi\pi} f(0,M_{\pi\pi}) 
  \nonumber \\
\times \frac{e^{(ik' - \frac{\beta}{2}) L}}{(ik' - \frac{\beta}{2})} 
\frac{e^{(i \Delta k L - \frac{\alpha - \beta}{2}) L} - 1}{(i \Delta k - \frac{\alpha - \beta}{2})} \label{f0}
\end{eqnarray}

For simplicity, we give here the probability for a fixed $M_{\pi\pi}$
(although the full calculations include the wide $\rho$):
\begin{eqnarray}
P_{\rho^0}(L,M_{\pi\pi})&=&4 \pi^2 |f(0)|^2 F^2 n^2 \nonumber \\
&\times\!&\!\frac{e^{-\alpha L}\!+\!e^{-\beta L}\!-\!2e^{-(\frac{\alpha+\beta}{2})L}\!\cos(\Delta k L)}
{(k'^2+\frac{\beta^2}{4})(\Delta k^2+ \frac{(\alpha-\beta)^2}{4})} \label{photoprodproba}
\end{eqnarray}
Eq. (\ref{photoprodproba}) illustrates some of the features of the system.
When $\Delta k L \gg 1$ the cosine fluctuates rapidly, leading to incoherent
$\rho$ production. However, when the coherence condition 
$\Delta k L \ll 1$ is fullfilled, the scattering amplitudes add in phase, and production is
coherent.

A $\rho$ inside a target is not directly observable. However, the way it interacts - 
electromagnetically, hadronically, or through $\rho$ conversion - is indirectly observable, by
studying the shower development.   The probability that an incident
photon interacts as a real $\rho$ is the integrated probability for finding a $\rho$, multiplied by the probability for
a $\rho$ to interact hadronically in length $dz$, or $dz/\beta$:
\begin{equation}
P(L) = \int_0^L \frac{dz |<\rho^0|\psi(z)>|^2}{\beta}.
\end{equation}
This equation is only properly normalized for $P(L)\ll 1$. The loss of photon intensity due to
the coherent $\rho$ reaction is not included.  For kaon regeneration, the analogous problem has been solved
recursively \cite{recursive}.  Here, we normalize the probabilities to sum to one in thick targets.  

The numerical values of $P(L)$ depend on $\sigma_{tot}(\rho A)$ and the cross section for a 
photon to interact hadonically (as a $\rho$), $\sigma_{hadr}(\gamma A)$.  They are related via 
\begin{equation}
\sigma_{hadr}(\gamma A) = F^2\sigma_{tot}(\rho A).  
\end{equation}
We determine these using two different methods: a Glauber calculation based on HERA data
on $\gamma p\rightarrow \rho p$, using a soft Pomeron model to
extrapolate the cross sections to higher energies, and a second calculation
that includes generalized vector meson dominance (GVMD) plus a component for direct
photon interactions \cite{engel}.  

We consider three materials: water, standard rock \cite{stdrock} 
and lead.  For water, we add the amplitudes for the cross sections for hydrogen and
oxygen.

The Glauber calculation uses the optical theorem \cite{blatt} and vector meson 
dominance model to
link the total $\rho p$ cross section to the differential $\gamma p$ cross section: 
\begin{eqnarray}
\sigma_{tot}(\rho p) = \frac{4 \sqrt{\pi}}{F} 
\sqrt{\frac{d\sigma(\gamma p \to \rho p)}{dt}|_{t=0}}
\end{eqnarray}
where $t$ is the squared momentum transfer. $d\sigma/dt(\gamma p\rightarrow\rho p$ comes from 
a fit to HERA data \cite{critenden}.

The cross sections for nuclear targets are found with 
a quantum Glauber calculation \cite{drell}: 
\begin{eqnarray}
\sigma_{tot}(\rho A) = 2 \int{d^2\vec{r} \bigg( 1 - e^{- \frac{\sigma_{tot}(\rho p)T_{A}(\vec{r})}{2} } \bigg) }
\end{eqnarray}
where $T_{A}(\vec{r})$ is the nuclear thickness function for a material with atomic number
$A$, calculated from a Woods-Saxon nuclear density, 
with a skin thickness of $0.53$ fm and density $\rho_0=1.16 A^{1/3} (1-1.16A^{-2/3})$.
\begin{figure}[t] 
\includegraphics[width=7.55cm]{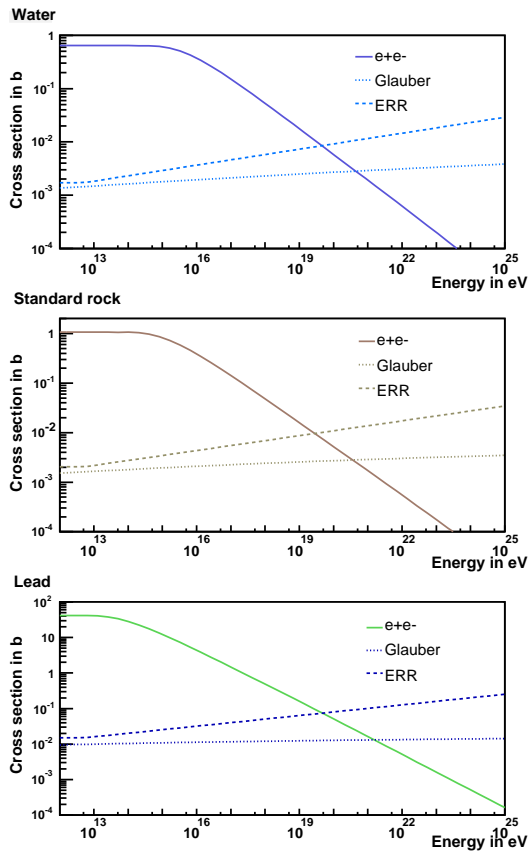} 
\caption{\label{photoncrosssection}The cross sections for photon interactions
to $e^+e^-$ pairs, and for incoherent photonuclear interactions in the Glauber and ERR models, for
water (top), standard rock (middle) and lead (bottom).} 
\end{figure}

The second approach follows
Engel, Ranft and Roessler (ERR) \cite{engel}.  It uses GVMD plus 
direct photon-quark interactions, to determine $\sigma_{hadr}(\gamma A)$.  ERR 
predict a steeper rise in the
cross section than the Glauber calculation.  We parameterize their results.
For $W_{\gamma p} < 10^{11}$ eV, the cross section is constant, while at higher energies
it rises as $W_{\gamma p}^{0.2}$.  The cross section scales as  
$A^{0.887}$, normalized so $\sigma(\gamma{\rm Pb})=15$ mb at low energies.
Again, $\sigma_{tot}(\rho A) = \sigma_{hadr}(\gamma A)/F^2$.
The ERR cross sections are similar to a  newer result that
computed photon cross sections using a dipole model \cite{strikman}.  

Figure \ref{photoncrosssection} compares the cross sections for 
$e^+e^-$ production and the two photonuclear models.
$\sigma_{ee}(\gamma A)$ is constant at low energies, but falls at higher energies
when the LPM effect becomes important \cite{lpm,eespencer}.
The hadronic cross sections rise slowly with energy. 
The hadronic curves agree at low energies, but diverge
as the energy rises.   In the ERR (Glauber) model, above $5\times10^{19}$ eV ($5\times10^{20}$ eV), photonuclear 
interactions predominate, and photons produce hadronic showers.  The cross-over
energy is almost material-independent for solids. 

\begin{figure}[tb] 
\includegraphics[width=6.3cm]{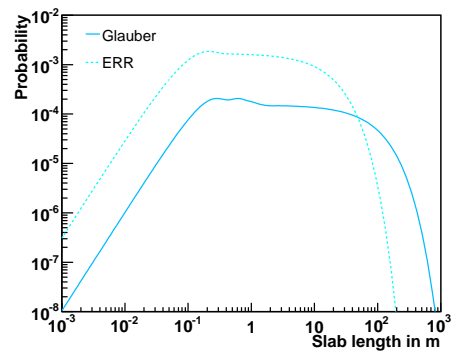} 
  \caption{\label{rhoprobability} $\rho$ probability as a function of depth in a slab 
for a $10^{23}$ eV photon incident on a water target.  The two broad peaks around 20 and 40 cm
correspond to $\Delta k L = 2n\pi$ at the $\rho$ pole mass.} 
\end{figure}

Figure \ref{rhoprobability} shows the probability of finding a $\rho^0$, Eq. (\ref{photoprodproba}),
as a function of depth in a thick target.
As with $K_s$ from kaon regeneration \cite{kaonregeneration}, the amplitudes add
and the probability initially rises
as the square of the depth.  After reaching a maximum, the probability decreases slowly as $\rho$
absorption competes with $\rho$ production, and then drops rapidly as the bare photons are
absorbed.

\begin{figure}[tb!] 
\includegraphics[width=5 cm]{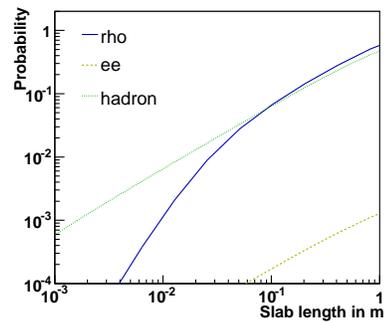} 
\caption{\label{figthickness} 
The probability of a $10^{24}$ eV photon interacting electromagnetically
(dashed blue line), in an incoherent hadronic interaction (dotted line), or as a 
coherently produced $\rho$ (solid line), as a function of target depth in a lead target.}
\end{figure}

Figure \ref{figthickness} shows the probability for a $10^{24}$ eV photon to interact
as a function of target thickness $L$.   For $L<10$ cm, the probability of a 
photon interacting as a coherently produced $\rho$ is quite small.
For $L>10$ cm, the probability is higher than for photonuclear
interactions and $e^+e^-$; it is the dominant interaction! 

\begin{figure*}[bt!] 
\includegraphics[width=12.5cm]{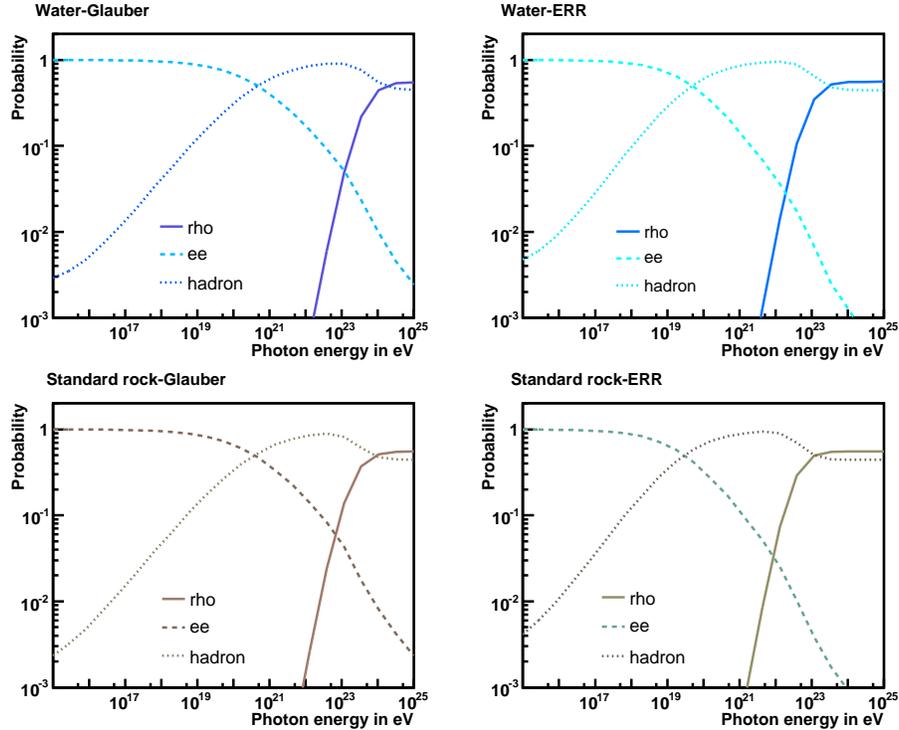} 
\caption{\label{thickslab} The normalized probabilities for a photon to interact as
an $e^+e^-$ pair, via incoherent hadronic interaction (`hadron'), or as an elastically produced
$\rho$ (`rho') in a 100 m thick water or standard rock target. This is thick enough for almost 
total absorption, so these results apply for an infinitely thick target.}
\end{figure*}

Figure \ref{thickslab} shows the probabilities for incident photons to interact electromagnetically 
($\gamma\rightarrow e^+e^-$ pair), via
incoherent hadronic interactions, and as coherently produced $\rho$, as a function of energy, in
a 100 m thick slab; in the displayed energy range, 100 m is effectively infinite thickness.
At energies above $10^{23}$ eV (for ERR)
to $10^{24}$ eV (for the Glauber calculation), coherently produced $\rho$ interactions dominate.
This dominance will reduce the interaction length of photons.  The reduction in shower length will 
have consequences for searches for ultra-high energy astrophysical 
$\nu_e$ \cite{radio,klein-erice}.

In conclusion, photons with energies above $10^{20}$ eV in solids or liquids are more likely to interact 
hadronically rather than
electromagnetically.  These hadronic showers are not subject to the LPM effect, and so are considerably 
more compact than purely electromagnetic calculations would predict.
At energies above $10^{23}$ eV, photons are most 
likely to interact by coherently
converting into a $\rho$, and then interacting hadronically in the target.  This is a new example of a coherent
process in bulk matter, similar to kaon regeneration or coherent neutrino scattering.
This coherent interaction further shortens the shower development.
The reduction in photon interaction length will shorten ultra-high energy electromagnetic showers, altering the
radio and acoustic emission.  It should be included in the models used in 
searches for ultra-high energy astrophysical $\nu_e$.

We thank Volker Koch for useful comments.  This work was funded by the U.S. National Science Foundation and the U.S.
Department of Energy under contract number DE-AC-76SF00098.

\end{document}